\documentstyle[preprint,prl,aps]{revtex}
\begin{document}
\draft

% for two column  activate the line below...                
%\twocolumn[\hsize\textwidth\columnwidth\hsize\csname@twocolumnfalse\endcsname

\title{
%LA-UR 96-3261\\
First-Order Melting of a Moving Vortex Lattice: Effects of Disorder}

\author{Daniel Dom\'{\i}nguez}
\address{Centro At\'{o}mico Bariloche, 8400 S. C. de Bariloche,
Rio Negro, Argentina}
\author{Niels Gr{\o}nbech-Jensen, and A. R. Bishop}
\address{Theoretical Division and Center for Nonlinear Studies, 
MS B262, Los Alamos National Laboratory,
Los Alamos, NM 87545}
\date{\today}
\maketitle
\begin{abstract}
We study the melting of a moving vortex lattice
through numerical simulations with the 
current driven 3D XY model with disorder.
We find that there is a first-order phase transition even
for large disorder when the corresponding 
equilibrium transition is continuous.
The low temperature phase is an anisotropic moving glass.
%The melting temperature of the driven system is 
%slightly lower than the equilibrium transition due to 
%current induced vortex loop unbinding.

\end{abstract}


% for two column  activate the line below...                            
%]                

\narrowtext

There is now growing experimental \cite{meltR,meltD,lopez,mag}
and theoretical \cite{meltth,hetzel,teitel,nos,edu,sudbo,others}
evidence that
the vortex lattice  (VL) in clean high $T_c$ superconductors melts
via a first order transition. Most of the experiments
consist of measurements of jumps in the linear  
resistance \cite{meltR,meltD,lopez}
and in the equilibrium magnetization \cite{mag}. These results are
consistent with  computer simulations in the 3D XY model
\cite{hetzel,teitel,nos,edu,sudbo} as well as 
in other models  \cite{others}. 
However, even in clean samples
there is always random weak pinning, which destroys the
crystalline order of the VL. 
Moreover, large disorder can
transform the first-order transition into a continuous
transition, as found both experimentally \cite{meltD} and
in simulations \cite{edu}.
Koshelev and Vinokur \cite{KV} have 
proposed that melting from a perfect lattice will 
only be possible for a rapidly {\it moving} VL. For driving
currents $I$ much larger than the critical current $I_c$, the 
VL is depinned and the effect of the
random potential in the moving VL is considerably weakened.
Recent experiments \cite{movexp1,movexp2,fendrich} 
in clean YBa$_2$Cu$_3$O$_7$ crystals
driven at currents  $I\gg I_c$ have indeed shown 
a melting transition of the moving VL.

On the other hand, Giamarchi and Le Doussal \cite{GLD} have shown that
the perturbation theory of \cite{KV} breaks down 
even at large vortex velocities,
because some modes of static disorder are still
present in the moving system. Balents and 
Fisher \cite{BF}, in the related system of sliding charge
density waves, found that the moving phase has 
only quasi-long range order in 3D. Therefore,
questions such us the existence and nature of a solid moving
phase and  which effects of the static disorder
remain once the VL is in motion are currently under active
discussion. 

In this Letter we study the behavior of 
a moving VL driven by currents $I\gg I_c$ as
a function of temperature and 
disorder.
We perform numerical simulations in a 3D current-driven XY model 
\cite{nos,edu}. 
A distinct advantage of this model with respect to 2D 
molecular dynamics simulations \cite{moon,ryu}, 
is that, besides the higher dimensionality,
the model allows for local fluctuations of the vorticity
(i.e. thermally nucleated vortex loops). 
Here, we find that the moving VL is a solid phase with
an anisotropic structure factor, consistent
with the theory of \cite{GLD},
which melts via a first-order phase transition.

The 3D XY Hamiltonian,
\begin{equation}
{\cal H}=-\sum_{{\bf r},\hat \mu} J_{\hat \mu} 
\cos\left[\Delta_{\hat\mu}\theta({\bf r})-
A_{\hat\mu}({\bf r})\right]\;,
\end{equation}
results from modeling the macroscopic properties
of a superconductor with the thermal fluctuations 
of the phase $\theta$
\cite{hetzel,teitel,nos,edu,sudbo}.
We consider  a network \cite{hetzel} given by
${\bf r}=r_1 \hat s_1 + r_2 \hat s_2 + r_3 \hat z$,
%  with $r_1, r_2, r_3$ integers, 
% and $\hat\mu=\hat n_1, \hat n_2, \hat n_3,\hat z$, 
with $\hat s_1=\hat x, \hat s_2=-\frac{1}{2}\hat x +\frac{\sqrt{3}}{2}
\hat y, \hat s_3 = \frac{1}{2}\hat x +\frac{\sqrt{3}}{2}\hat y$.
The phase difference is 
$\Delta_{\hat\mu}\theta({\bf r})=\theta({\bf r}+\hat\mu)-\theta({\bf r})$.
The gauge factor $A_{\hat\mu}({\bf r})=\frac{2\pi}{\Phi_0}\int_{\bf r}^{
{\bf r}+\hat\mu}{\bf A}\cdot d{\bf l}$ 
($\Phi_0=h/2e$) depends on the magnetic
induction ${\bf B}=\nabla\times{\bf A}=B\hat z$.
(We neglect fluctuations in ${\bf B}$, valid for 
$(\Phi_0/B)^{1/2}\ll\lambda$).
We have $A_z(r)=0$ and  
$\sum_{xy} 
A_{\hat\mu}({\bf r})=2\pi f =2\pi Bs^2\sqrt{3}/4\Phi_0$,
with the network discretization $s=|\hat\mu|$.
We model disorder  with a 
random  $J_{\hat\mu}$  in the interval
$[(1-\delta/2)J,(1+\delta/2)J]$ and
$\langle J_{\hat\mu}\rangle=J(T)=\Phi_0^2s/16\pi^3\lambda^2(T)$ 
with $\lambda(T)$ the 
mean-field penetration depth.
We consider a field of $f=1/6$ and a system
of size $L^3$. This model was studied in Ref.~\cite{hetzel}, 
where it was shown that the VL melting
transition is first order. 
It also has a simultaneous loss of superconducting
coherence both in the $xy$ planes and the $z$ direction \cite{nos}, as seen
experimentally in YBa$_2$Cu$_3$O$_7$ \cite{lopez}.

The current $I_{\hat\mu}({\bf r})$  
in each  bond of the network is \cite{nos}
\begin{equation}
I_{\hat\mu}({\bf r})=\frac{\Phi_{0}}{2\pi{R}_N}
\frac{d \Delta_{\hat\mu}\theta({\bf r})}
{dt} + I_{0,\hat\mu} \sin\left[\Delta_{\hat\mu}\theta({\bf r})-
A_{\hat\mu}({\bf r})\right] +
\eta_{\hat\mu}({\bf r},t)
\end{equation}
with $I_{0,\hat\mu}=2\pi J_{\hat\mu}/\Phi_0$, and 
${R}_N$ the shunt resistance. 
The thermal noise term is taken to have correlations 
$\langle \eta_{\hat\mu}({\bf r},t) \eta_{\hat\mu'}({\bf r'},t')\rangle=
(2k_BT/{R}_N) \delta_{\hat\mu,\hat\mu'}
\delta_{{\bf r},{\bf r'}}\delta(t-t')$.
Together with the condition of current conservation,
\begin{equation}
\sum_{\hat\mu} [I_{\hat\mu}({\bf r})-I_{\hat\mu}({\bf r}-\hat \mu)]
=\Delta_{\hat\mu}\cdot I_{\hat\mu}({\bf r})=I_{ext}({\bf r}),
\end{equation}
this determines the full set of dynamical equations. The boundary conditions
are periodic along the $\hat x$  ($\hat s_1$) and $\hat z$ directions, 
and  open  in the $\hat s_2$ direction,
with a current bias $I$  corresponding to
$I_{ext}({\bf r})=I(\delta_{r_2,0}-\delta_{r_2,L})$.
We simulate this set of  equations
with the same numerical methods and integration parameters 
as in Ref.~\cite{nos}.

We calculate the normalized voltage drop  along the direction of the current
as $v=\frac{\tau_J}{L^3} \sum_{r_1,r_3}
\left\langle
{\dot\theta(r_1,L,r_3)-\dot\theta(r_1,0,r_3)}
\right\rangle$. 
In Fig.~1(a) we show the dc resistance $R=v/i$ 
for a high current $i=0.2$ (currents are normalized 
by $\langle I_{0,\hat\mu}\rangle$) as 
a function of increasing and decreasing temperature,
for systems with disorder $\delta=0.01,0.1,0.8$.
The driving current is well above 
the critical current $i_c(T=0)\approx 0.04$, so the
moving lattice is depinned. 
The local vorticity along the direction $\hat\nu$
is given by  
$n_{\hat\nu}({\bf R},t)=-\sum_{\rm plaquette}\,
{\rm nint}\left[({\Delta_{\hat\mu}\theta({\bf r})
-A_{\hat\mu}({\bf r}))/{2\pi}}\right]$ 
for $\hat\nu\perp\hat\mu$; with ${\rm nint}[x]$ the nearest integer to $x$.
Vorticity is conserved both locally: $\Delta_\mu\cdot n_{\hat\mu}({\bf R})=0$,
and globally: $\langle n_z(R,t)\rangle=f$, 
$\langle n_{\hat s_i}(R,t)\rangle=0$.
In order to study vortex lattice melting, we
calculate the vortex structure function: 
$
S({\bf k})=\frac{1}{L^3}\sum_{r_3}\langle n_z({\bf k},r_3)
n_z(-{\bf k},r_3)\rangle
$.
In Fig.1(b) we show the intensity of  one
of the Bragg peaks of the VL,
$S_G=S({\bf G}_1)$, with ${\bf G}_1=(\frac{2\pi}{3},
\frac{2\pi}{\sqrt{3}})$
as a function of $T$ for $\delta=0.01,0.1,0.8$.
Besides the field-induced vortex lines, it is also
possible to have thermally induced vortex loops.
Therefore, we calculate the average number of extra vortices along
the $z$ direction, $n_z=\langle |n_z(R,t)|\rangle -f$, 
and the average 
vortex excitations in the $xy$ directions $n_{xy}=
\langle |n_{xy}(R,t)| \rangle$.
This is shown in Fig.1(c) as a function of $T$
for $\delta=0.1$ only.

The results of Fig.1(b) show that the moving VL has a first-order
melting transition where $S_G$ vanishes
with a sudden drop of nearly two orders of magnitude at a melting
temperature $T_M^I$. We also see hysteresis
when increasing and decreasing $T$. We find that for weak disorder
$\delta < \delta_c\sim 0.5$ the melting transition does
not show any significant variation with disorder 
(in Fig.1(b) the
results for $\delta=0.01$ and $\delta=0.1$ are very similar).
This means that the effect of random pinning in 
the moving VL is negligible for small disorder.
For $\delta>\delta_c$, $T_M^I$ shifts to
lower temperatures and the hysteresis loop slightly increases
(shown for $\delta=0.8$ in Fig.1).
A finite size study of the transition is shown in Fig.~2 for
$\delta=0.1$, for system sizes $L=12,18,24$ \cite{periodic}.
We see that for $L=12$, $S_G$ vs. $T$ shows a broad transition with little
hysteresis. After increasing the system size, 
there is a clear hysteretic  jump in
$S_G$, which becomes sharper when going from $L=18$ to $L=24$, suggesting
that there is a first-order transition in the thermodynamic limit.
The melting temperature $T_M^I$ slightly decreases when going
from $L=18$ to $L=24$, meaning that the asymptotic behavior has not been
reached yet. (However, the difference is close to the temperature
resolution of this Langevin simulation, $\Delta T \approx 0.01$).
 We also find that the superconducting coherence along the
$z$ direction vanishes simultaneously with the melting transition,
from the calculation of the helicity modulus $\Upsilon_z$ (not shown
here), similar to what we found in equilibrium
for $\delta=0$ \cite{nos}. 
%We have found this first-order single
%transition for any value of the disorder  for the {\it moving} VL
%(this is in contrast to the effect of disorder in the {\it static} VL 
%where large disorder can split the single transition into 
%two transitions \cite{edu}).

The experiments of Ref.~\cite{movexp1,movexp2} consisted
of measurements of voltage and differential resistance for
the VL driven at $I\gg I_c$.
A small dip in the differential
resistance was observed at temperatures close to but below
the equilibrium $T_M$. In Fig.1(a) we see
that the dc voltage for $T<T_M^I$ is approximately
constant; it has a small dip at $T_M^I$ and then  rises
with temperature for $T>T_M^I$. The dip in the voltage
is very small for weak disorder (within the statistical error),
and increases with disorder, being very noticeable for $\delta>\delta_c$.
The voltage dip is usually associated with a peak effect in
the critical current happening right below  melting 
\cite{movexp1,movexp2}. From our simulation results, we see
that this is only a weak evidence for a melting transition.
Recent magnetization measurements 
done by Fendrich {\it et al} \cite{fendrich}
give a better signature of the melting of the moving VL.

It is interesting to study the behavior of thermally excited vortices.
In Fig.1(c) we show the number
of vortex excitations in the directions parallel to the field,
$n_z$, parallel to the current, $n_y$, and
perpendicular to both the field and the current, $n_x$. We
see that all of them increase with $T$ (at low $T$ it
is a thermally activated process $n_\nu \sim \exp(-U_\nu/T)$),
and they have a sudden jump and hysteresis at $T_M^I$. 
We also find that $n_x>n_y\gg n_z$ for all $T$. When compared with
their values in equilibrium  we find that $n_x(I)>n_x(0)$,
$n_y(I)<n_y(0)$ and $n_z(I)>n_z(0)$ (near equilibrium
$n_x(0)=n_y(0)\gg n_z(0)$). This means that the effect
of the drive is to orient the vortex loops in the
plane perpendicular to the current, and to increase the
average size of the loops.  Also the relative fraction
of vortex excitations in the field direction $n_z/(n_x+n_y)$ increases
with $I$, thus increasing the entropy of the moving phase.
These effects tend to lower the temperature
for loop unbinding with respect
to the melting and unbinding temperature $T_M^0$ of the static VL.

Let us now analyze the structure of the moving VL.
Surface plots of $S({\bf k})$ are shown in  Figs.3(a-b)
for temperatures below and above $T_M^I$.
We find that below $T_M^I$ there are Bragg peaks which suddenly disappear 
at $T_M^I$. The Bragg peak structure of the low temperature moving
phase is very anisotropic as can be seen in Fig.3(a).
There are two possible non-degenerate orientations of the 
VL relative to the external current. In the 2D simulations of
Moon {\it et al.} \cite{moon} the VL is oriented with one of 
the reciprocal lattice
vectors perpendicular to the driving Lorentz force, while here one
is parallel to it. This orientation is favored here because of the 
network discretization relative to the direction of the
applied current. (In the other orientation the VL is frustrated.)
In Fig.3(a) we see that
the Bragg peak with reciprocal lattice vector ${\bf G}_0=\pm 
(\frac{4\pi}{3},0)$, 
parallel to the direction of motion, is considerably smaller
than the peaks at ${\bf G}_1=\pm (\frac{2\pi}{3},\frac{2\pi}{\sqrt{3}})$  and
${\bf G}_2=\pm (-\frac{2\pi}{3},\frac{2\pi}{\sqrt{3}})$. 
A similar result was found
in \cite{moon} for the
peaks with components parallel to the direction of motion. However, when
increasing system sizes, we do not see here any significant
decrease on the height of the ${\bf G}_0$ peak, as reported in \cite{moon}. 
In Fig.3(c-d) we analyze in detail the finite size scaling behavior
of the peak at ${\bf G_2}$.  
The peak is very anisotropic:
it  has a finite width in the direction perpendicular to vortex
motion [Fig.3(c)] whereas it has almost zero width in the direction 
parallel to vortex motion [Fig.3(d)].  
In Fig.3(c) we show a finite size scaling analysis of the peak width
along the $y$ direction : $S(\delta k_y,L)\sim L^{2-\nu}F(\delta k_y
L)$ for $L=12,18,24$, with $\delta k_y=k_y-G_{2y}$. 
Due to the discreteness of the 
network there are few points  to consider, and the scaling results
are not very precise. We obtain $\nu \sim 0.15 \pm 0.1$. We find
a similar result for the other peaks, both for ${\bf G}_1$ and for
${\bf G}_0$,  but with a greater inaccuracy  in the latter case.
Despite the large uncertainty, from our results we can be sure
that $\nu < 0.3$. In the 2D simulations of \cite{moon} an exponent
of $\nu_{2D} \approx 0.53$ was obtained, with seemingly isotropic Bragg
peaks. The anisotropic peaks found here are more consistent with
the discussion of Giamarchi and Le Doussal in \cite{GLD}, where it was
shown that the effects of the static disorder are still relevant
in the direction transverse to vortex motion. The fact that
in our case the low temperature moving phase seems more ``ordered''
(i.e. smaller $\nu$) can be due to: (a) three-dimensionality; 
(b) the long-range (logarithmic) interactions between vortices. 
In any case, our results are consistent with the recent experiment
of Fendrich {\it et al.} \cite{fendrich} where a large external
drive does not affect the first-order character of the melting
transition, indicating that there is a low temperature moving solid.

We have made a systematic comparison of our results for the moving
VL ($I\gg I_c$) with simulations near equilibrium ($I\ll I_c$) for
different values of $\delta$. The main result
is summarized in the phase diagram shown in Fig.4. 
We see that for increasing disorder the equilibrium 
first-order transition turns into a continuous transition
or a crossover for $\delta\gtrsim 0.5$, in agreement with
the results of Jagla and Balseiro \cite{edu}. On the contrary,
the phase transition of the moving VL {\it is always of first-order
character}, the only effect of large disorder $\delta\gtrsim 0.5$ 
is in lowering $T_M^I$. Also we find that $T_M^I < T_M^0$ for any
value of the disorder. This is due to the coincidence of
melting with vortex loop unbinding: since the bias current 
tends to unbind loops more easily, $T_M^I$ should be lower
than $T_M^0$, as we discussed above. Also
at zero field, where the phase transition is driven
by vortex loop unbinding, a bias
current considerably lowers the critical temperature
\cite{nos2}.  Recently, it has been suggested
by Nguyen {\it et al.} \cite{sudbo} that the presence
of vortex loops in the melting transition could explain
the anomalous behavior of the entropy jump found by
Zeldov {\it et al.} \cite{mag}. 
From our results, Fig.4, we
believe that a measurement of a finite $\Delta T_M=T_M^0-T_M^I$
could be an experimental probe of the role of vortex
loops in the melting transition.
Also a ``shaking temperature'' effect could give $T_M^I<T_M^0$ \cite{KV}.
However, this gives a $\Delta T_M$ which depends on disorder,
and should vanish for $\delta\rightarrow0$. 
In the experiment of Safar {\it et al.}
\cite{movexp1}, with a field $B\parallel c$, a dip in the 
differential resistance was observed very close to,
but below, the equilibrium melting transition. The experiment
of Fendrich {\it et al.} \cite{fendrich} does not show a reasonable
dependence of $T_M$  with the bias current.  This can be due to:
(a) the $\Delta T_M$ is too small to be discernible within
experimental resolution, 
or (b) the bias current of \cite{fendrich} is not high
enough (here, the current is such that
the VL is in a flux flow state at $T=0$, whereas in \cite{fendrich}
there is no dissipation at low $T$).  
In the experiment of D'Anna  {\it et al.} \cite{movexp2},
with  $B\parallel ab$ a dip in the differential resistance
appears clearly below the equilibrium $T_M$.
Since it is easier to have thermal nucleation of loops parallel 
to the $ab$ planes,
this could lead to a measurable $\Delta T_M$ for $B\parallel ab$.

In conclusion, we find that the low temperature moving phase is an
anisotropic glass which melts via a first-order transition
in 3D, even for large values of disorder. Our results
stress the three-dimensionality of the VL: 
there is  a more ``ordered'' moving
phase when compared with 2D simulations \cite{moon}, and at melting there
is a simultaneous loss of superconducting coherence along the $z$ axis.

D. D. acknowledges discussions with H. Safar.
This work was supported by the U.S.D.O.E..

\begin{figure}
\caption{(a) DC voltage over current
$R=v/i$ as a function of temperature $T$ for a bias current $i=0.2 \gg
i_c$ and $L=18$.
Dotted line: $\delta=0.01$, triangles: $\delta=0.1$; squares:
$\delta=0.8$. The melting temperatures in equilibrium, $T_M^0$, 
and for the driven system, $T_M^I$, are indicated.
(b) Intensity of the  Bragg peak
$S_G$ vs. $T$. 
(c) Plot of the density of thermally excited vortices $n_\nu$ vs. 
$T$, for $\delta=0.1$.}
\end{figure}

\begin{figure}
\caption{Plot of the intensity of the  Bragg peak
$S_G$ vs. $T$ for $\delta=0.1$ and different system sizes.}
\end{figure}

\begin{figure}
\caption{Surface plots of the structure factor $S({\bf k})$ 
for $\delta=0.1$, current $i=0.2$ and $L=18$. 
For (a) $T=1.0$, (b) $T=1.2$. 
(c) Finite size scaling plot of the k-dependence of the Bragg peak in the
direction transverse to vortex motion. (d) Same as (c)
but in the direction parallel to vortex motion. Results
for $\delta=0.1$ and $T=0.9$.}
\end{figure}

\begin{figure}
\caption{Disorder-temperature phase diagram. Triangles:
equilibrium transition temperature $T_M^0$. Squares:
transition temperature of the driven system $T_M^I$.
Continuous lines: first-order transitions.
Dashed line: continuous transition. 
Dotted lines: limit of the hysteresis loops
for a given cooling rate. Results for $L=18$.}
\end{figure}

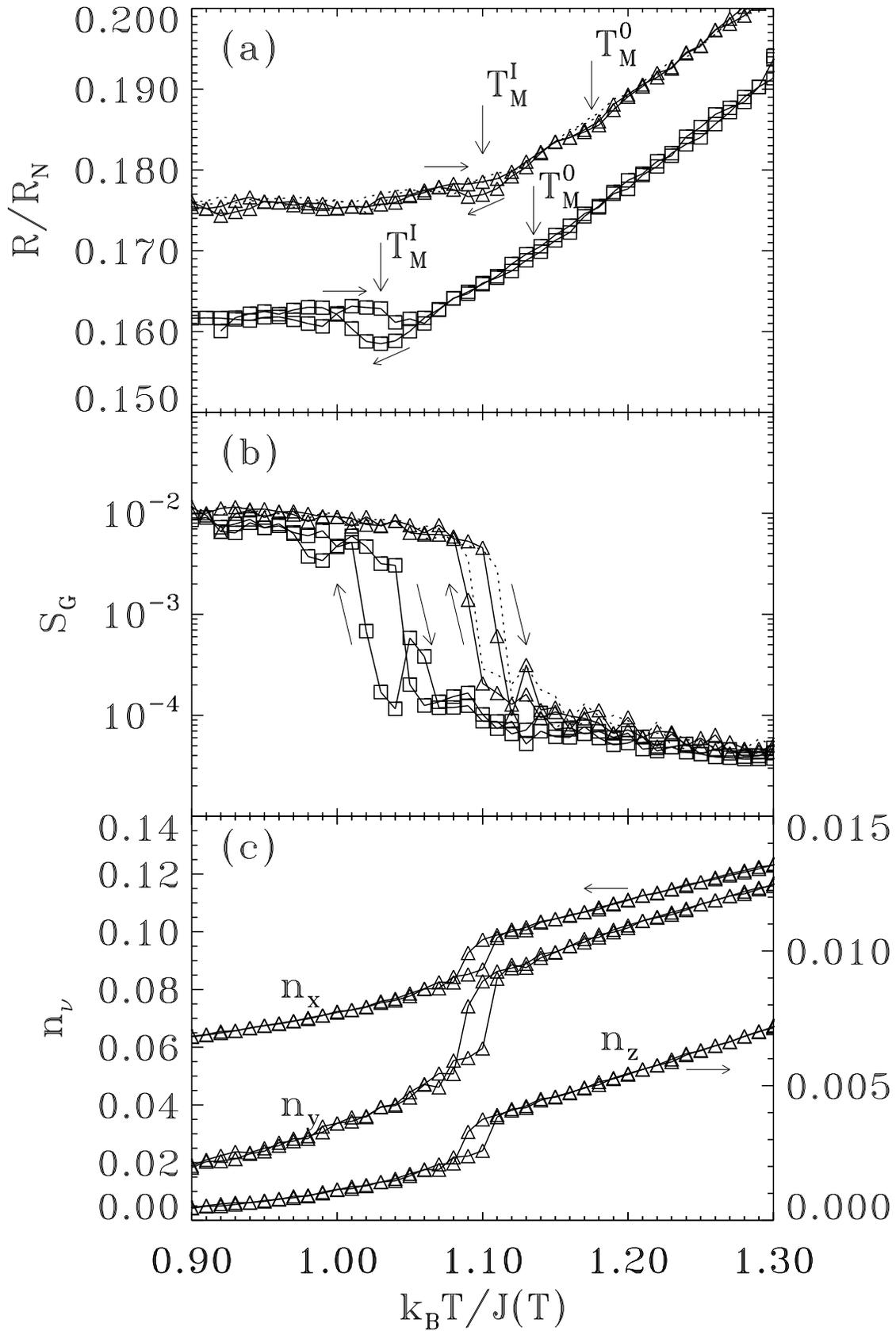
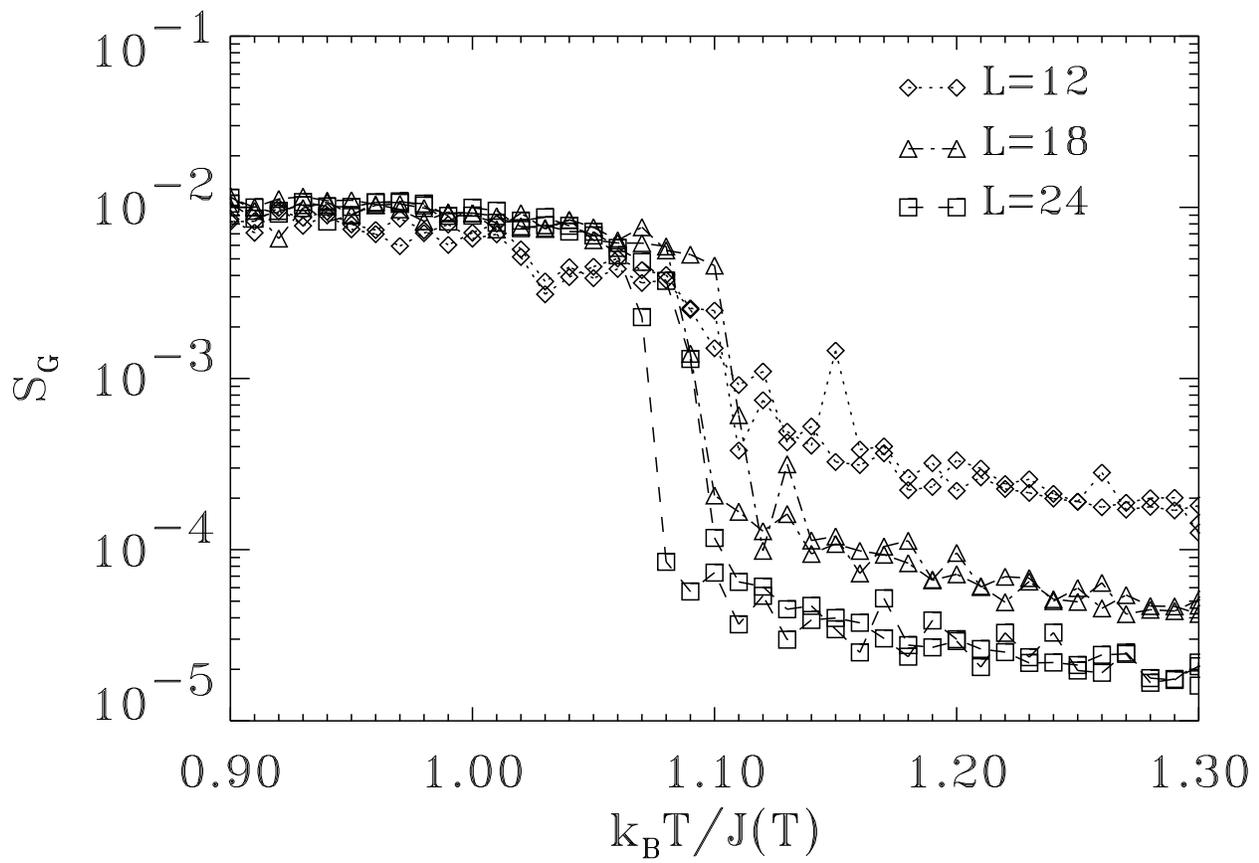
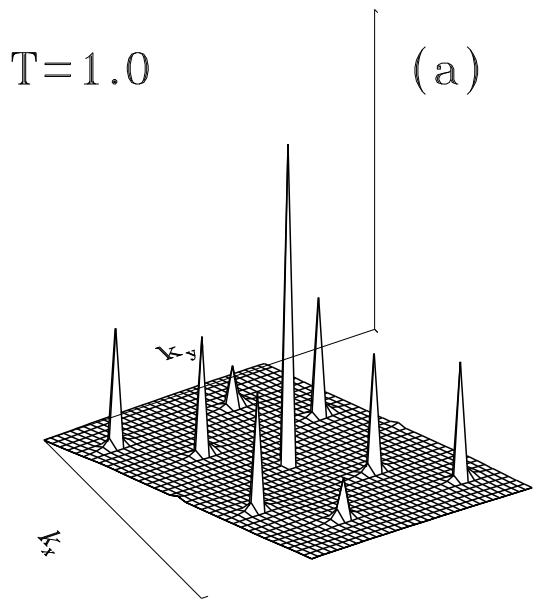
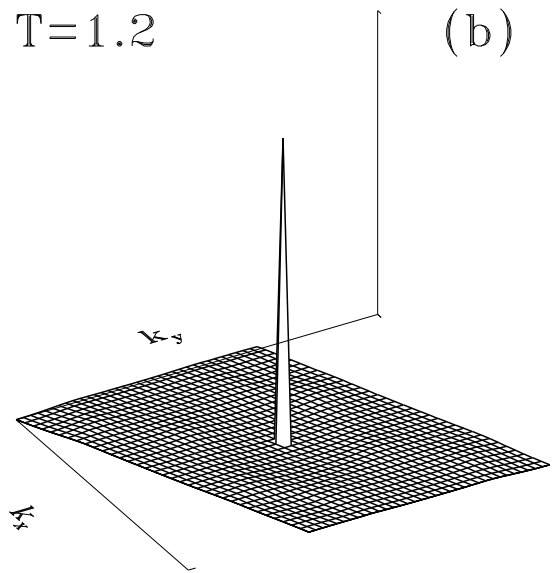
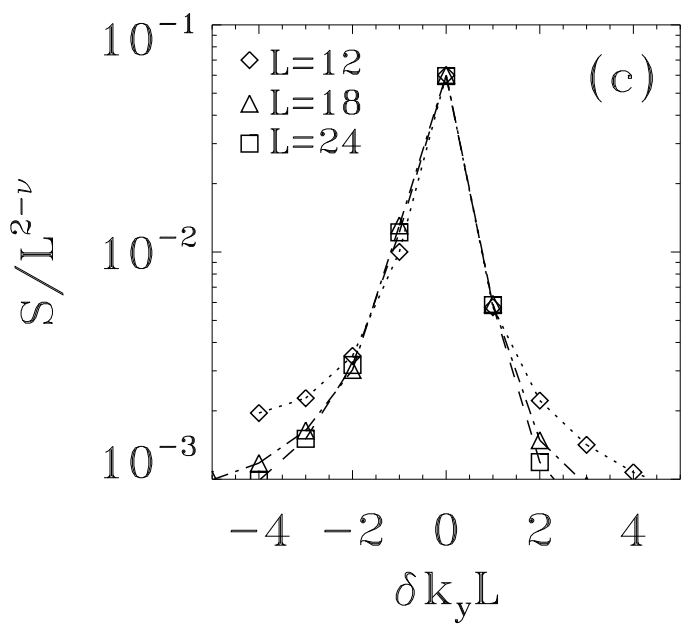
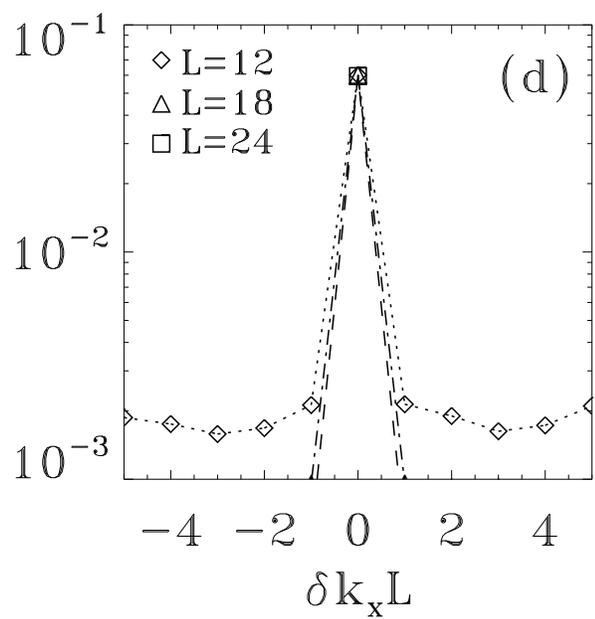
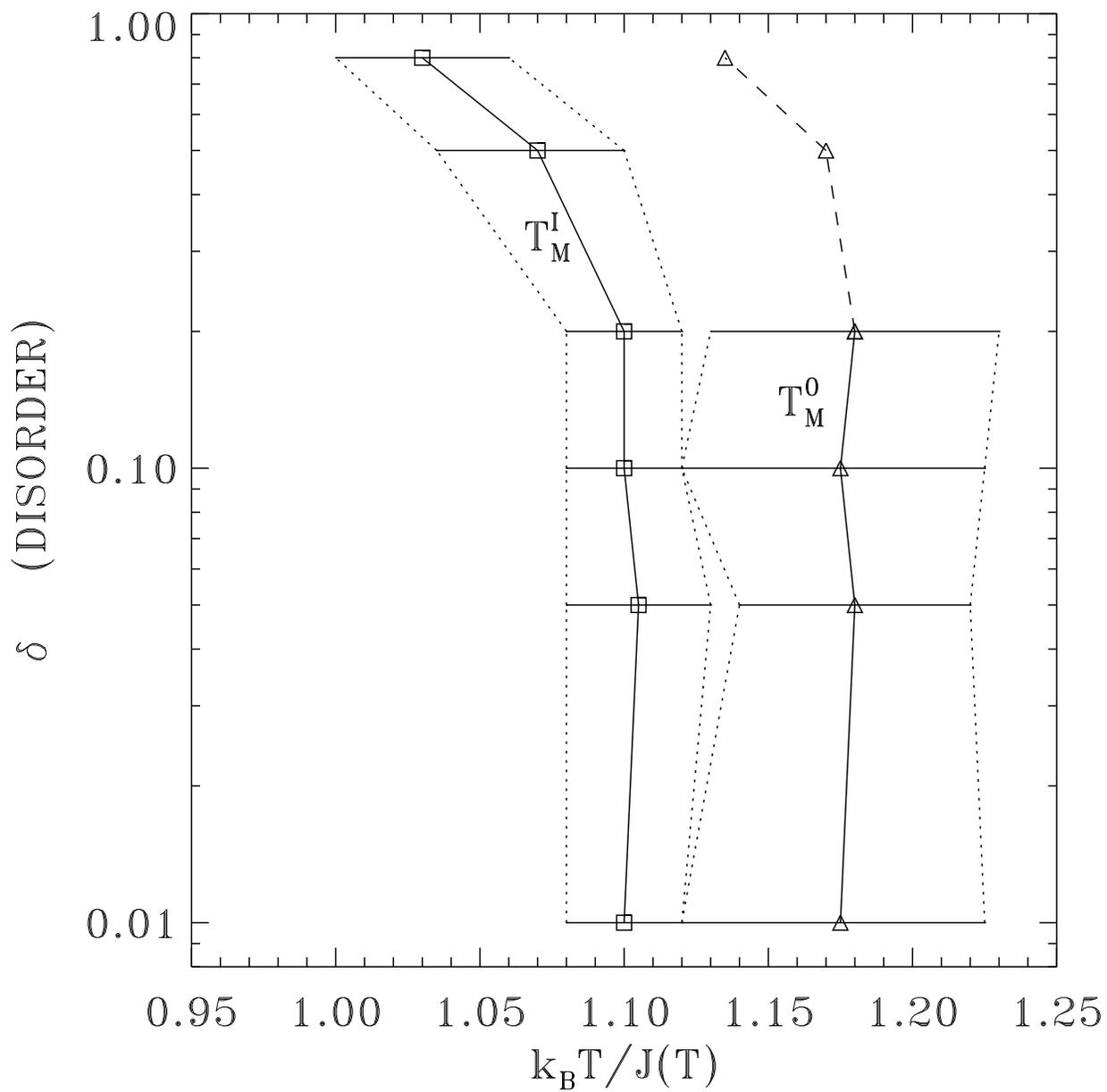
\end{document}